\documentclass[%
 reprint, 
 prc,
 amsmath,amssymb,
 aps,
floatfix,
superscriptaddress
]{revtex4-1}

\usepackage{graphicx}
\usepackage{dcolumn}
\usepackage{bm}

\begin{document}

\title{Measurement of the $^{18}$Ne($\alpha,p$)$^{21}$Na reaction with ANASEN at \textbf{$E_{c.m.} = $} 2.5 - 4 MeV}

\author{M.\ Anastasiou}
 \email{anastasiou2@llnl.gov}
\affiliation{Physics Department, Florida State University, Tallahassee, Florida 32306, USA}
\affiliation{Lawrence Livermore National Laboratory, Livermore, California 94550, USA}
\author{I.\ Wiedenh\"over}%
 \email{iwiedenhoever@fsu.edu}
\affiliation{Physics Department, Florida State University, Tallahassee, Florida 32306, USA}
\author{J. C. Blackmon}
\affiliation{Department of Physics and Astronomy, Louisiana State University, Baton Rouge, Louisiana 70803, USA}
\author{L. T. Baby}
\affiliation{Physics Department, Florida State University, Tallahassee, Florida 32306, USA}
\author{D. D.~Caussyn}
\affiliation{Physics Department, Florida State University, Tallahassee, Florida 32306, USA}
\author{A. A.~Hood}
\affiliation{Cyclotron Institute, Texas A\&M University, College Station, Texas 77843, USA}
\affiliation{Department of Physics and Astronomy, Louisiana State University, Baton Rouge, Louisiana 70803, USA}
\author{E. Koshchiy}
\affiliation{Cyclotron Institute, Texas A\&M University, College Station, Texas 77843, USA}
\author{J. C. Lighthall}
\affiliation{Department of Physics and Astronomy, Louisiana State University, Baton Rouge, Louisiana 70803, USA}
\author{K. T. Macon}\affiliation{Department of Physics and Astronomy, Louisiana State University, Baton Rouge, Louisiana 70803, USA}
\author{J. J. Parker}
\affiliation{Physics Department, Florida State University, Tallahassee, Florida 32306, USA}
\author{T. Rauscher}
\affiliation{Department of Physics, University of Basel, 4056 Basel, Switzerland}
\affiliation{Centre for Astrophysics Research, University of Hertfordshire, Hatfield AL10 9AB, United Kingdom}
\author{N. Rijal}
\affiliation{Physics Department, Florida State University, Tallahassee, Florida 32306, USA}
\affiliation{Facility for Rare Isotope Beams, Michigan State University, East Lansing, Michigan 48824, USA}

\date{\today}

\begin{abstract}
The $^{18}$Ne($\alpha,p$)$^{21}$Na reaction plays a significant role in Type-I X-ray bursts. It is a major path in the breakout from the hot-CNO cycles to the synthesis of heavier elements in the $\alpha p$-- and $rp$-processes. An experiment to determine the cross section of this reaction was performed with the 
ANASEN active-target detector system, determining the cross section at energies between 2.5 and 4 MeV in the center-of-mass frame. The measured cross sections for reactions populating the ground state in $^{21}$Na are consistent with results obtained from the time-inverse reaction, but significantly lower than the previously published experimental data of direct measurements. The total cross sections are also compared with those derived from indirect methods and statistical-model calculations. This experiment establishes a new experimental data set on the excitation function of the $^{18}$Ne($\alpha,p$)$^{21}$Na reaction, revealing the significance of the excited states' contributions to the total reaction cross section and allowing to separate the contribution of the $(\alpha,2p)$ reaction. The impact of the measured cross section on thermal reaction rates is discussed. 
  
\end{abstract}

\pacs{Valid PACS appear here}
\maketitle


\section{\label{sec:level1-Intro}Introduction}
Type-I X-ray bursts occur when hydrogen- and helium-rich matter is accreted onto a neutron star in a close-binary system \cite{Book,Schatz06}. 
In this environment, hydrogen is converted into helium via the hot-CNO (HCNO) cycles, in which the energy release is limited by the $\beta$-decay rates of the $^{14}$O ($t_{1/2} =$ 70.6~sec), $^{15}$O ($t_{1/2} =$ 122~sec) and $^{18}$Ne ($t_{1/2} =$ 1.67~sec) isotopes, and thereby is mostly independent of temperature. The temperature-dependent energy production, observed in explosive hydrogen burning, can only occur after the breakout from the closed HCNO cycles and is initiated through the activation of the $\mathrm{^{15}O(\alpha,\gamma)^{19}Ne}$ and $^{18}$Ne($\alpha,p$)$^{21}$Na reactions \cite{rpProcess,Wiescher_1999}. Therefore, these reactions carry a large impact on the timescale and onset conditions for the explosion. The role of individual nuclear reactions on the properties of X-ray burst events has been studied in significant detail, e.g. by Parikh {\it et al.} \cite{Parikh08,Parikh13}, and Cyburt {\it et al.} \cite{Cyburt2016}. The $^{18}$Ne($\alpha,p$)$^{21}$Na reaction has been identified as one of the most important reactions in its impact on the light curve and nucleosynthesis in X-ray burst events. 

The available experimental information on the $^{18}$Ne($\alpha,p$)$^{21}$Na reaction is limited and somewhat contradictory. Direct measurements have remained difficult, as they require cross section measurements significantly below 10 mbarn with beams of a T$_z$=-1 short-lived isotope at low energies, combined with the use of a gas-target to provide the $\alpha$ particles. The only experimental data on the $(\alpha,p)$ reaction was reported in Bradfield-Smith {\it et al.} \cite{bradfield} and Groombridge {\it et al.} \cite{groom}, which presented results from experiments performed at the Louvain-la-Neuve laboratory. Both utilized a $^{18}$Ne beam at reaction energies between 1.7 and 3~MeV in the center-of-mass system and an experimental setup of a gas-filled chamber with silicon detectors. These experimental results were published as a set of resonance parameters, which were used to deduce thermal reaction rates. 

Experiments on the time-inverse reaction $^{21}$Na($p,\alpha)^{18}$Ne by Sinha {\it et al.} \cite{ANL} and Salter {\it et al.} \cite{salter} have reached a higher sensitivity towards lower reaction energies than the time-forward studies, owing to the higher-intensity $^{21}$Na beam and the use of solid polyethylene targets. However, their interpretation is complicated by the fact that the $^{18}$Ne($\alpha,p$)$^{21}$Na reaction may proceed to excited $^{21}$Na states even at the lowest reaction energies, which cannot be examined by the time-inverse reaction with the $^{21}$Na ground state. It was noted in Salter {\it et al.} \cite{salter} that their results point to much lower cross sections than Groombridge {\it et al.} \cite{groom} reported.

Additional experimental information to determine the astrophysical reaction rates is available in the form of the resonance parameters in the $^{22}$Mg compound system, which determine the astrophysical $^{18}$Ne($\alpha,p$)$^{21}$Na reaction rate. 
Examples of efforts to characterize the resonance properties include experiments with the $^{24}$Mg($p$,$t$)$^{22}$Mg reaction by Matic {\it et al.} \cite{Matic09} and Chae {\it et al.} \cite{Chae09}, and more recently with the resonant proton scattering $^{21}$Na($p,p$)$^{21}$Na reaction in inverse kinematics by He {\it et al.} \cite{jjhe2013}, Zhang {\it et al.} \cite{zhang2014} and Ruiz {\it et al.} \cite{ruiz2005}. 

A comprehensive analysis of the available experimental information on the $^{22}$Mg compound system was performed by Mohr {\it et al.} \cite{mohr2013, mohr2014}. 
In the narrow, non-interacting resonance approximation, the stellar reaction rate $N_{A}<\sigma v>$ of the $^{18}$Ne($\alpha,p$)$^{21}$Na reaction depends on the sum of the contributing resonances: 
\begin{equation}
    N_{A}<\sigma v> \sim \sum_{i} (\omega\gamma)_{i} e^{-E_i/kT_9},
\end{equation}
with resonance energies $E_i$ and resonance strengths $(\omega\gamma)_i$. The resonance strength for a resonance with spin $J$ for the $^{18}$Ne($\alpha,p$)$^{21}$Na reaction is determined by
\begin{equation}
    \omega\gamma_{\alpha p} = (2J+1)\frac{\Gamma_{\alpha}\Gamma_p}{\Gamma} \approx (2J+1)\Gamma_{\alpha},
\end{equation}
where the approximations $\Gamma_p \approx \Gamma$ and $\Gamma_{\alpha} \ll \Gamma_p$ are used. 
The analysis by Mohr {\it et al.} \cite{mohr2013, mohr2014}, employs the total widths $\Gamma$ and the excitation energies of the $^{22}$Mg compound nucleus, which define the resonance energies $E_i$ in the $^{18}$Ne($\alpha,p$)$^{21}$Na reaction, from the $^{24}$Mg($p,t$)$^{22}$Mg transfer experiment \cite{Matic09}. The $J^{\pi}$ assignments were obtained initially from Ref.\ \cite{Matic09, Chae09} and confirmed from the latest experiments on resonant proton scattering  \cite{jjhe2013, zhang2014}. However, the transfer reactions cannot provide the required $\Gamma_{\alpha}$ widths, and thus the resonance strengths $\omega\gamma_{\alpha p}$. As a result, these have to be determined from theory or obtained from direct experiments. Since the partial widths $\Gamma_{\alpha}$ of $^{22}$Mg levels in the relevant energy window have not been measured directly yet,  they were estimated using the assumption of mirror symmetry in the wave functions of corresponding $^{22}$Mg and $^{22}$Ne levels in Ref.\ \cite{mohr2014}. For levels with no spectroscopic information, the partial widths $\Gamma_{\alpha}$ were calculated from a Porter-Thomas distribution following Ref.\ \cite{porter}. Additionally, while the time-reverse $^{21}$Na($p,\alpha$)$^{18}$Ne reaction information \cite{salter, ANL} was used in the Mohr {\it {et al.}} studies \cite{mohr2013, mohr2014}, to determine a lower limit for the time-forward reaction cross section, the results of  Groombridge {\it et al.} \cite{groom} were excluded from the study due to the large disagreement with the time-reverse-reaction data. Both studies of Ref.\ \cite{mohr2013} and \cite{mohr2014} arrived at cross sections and reaction rates significantly lower than the ones determined by Groombridge {\it et al.} \cite{groom}. 
 
This work presents the results of a new experiment to measure the excitation function of the $^{18}$Ne($\alpha,p$)$^{21}$Na reaction at energies relevant to X-ray burst events at peak temperatures ($T\sim 2$ GK), and to clarify the discrepancies between the direct experimental data and the information obtained from indirect methods. 

\section{\label{sec:experiment}Experiment and Calibrations}

\begin{figure*}[t]
	\centering
	\begin{minipage}{0.8\textwidth}
		\centering
		\includegraphics[width=\textwidth]{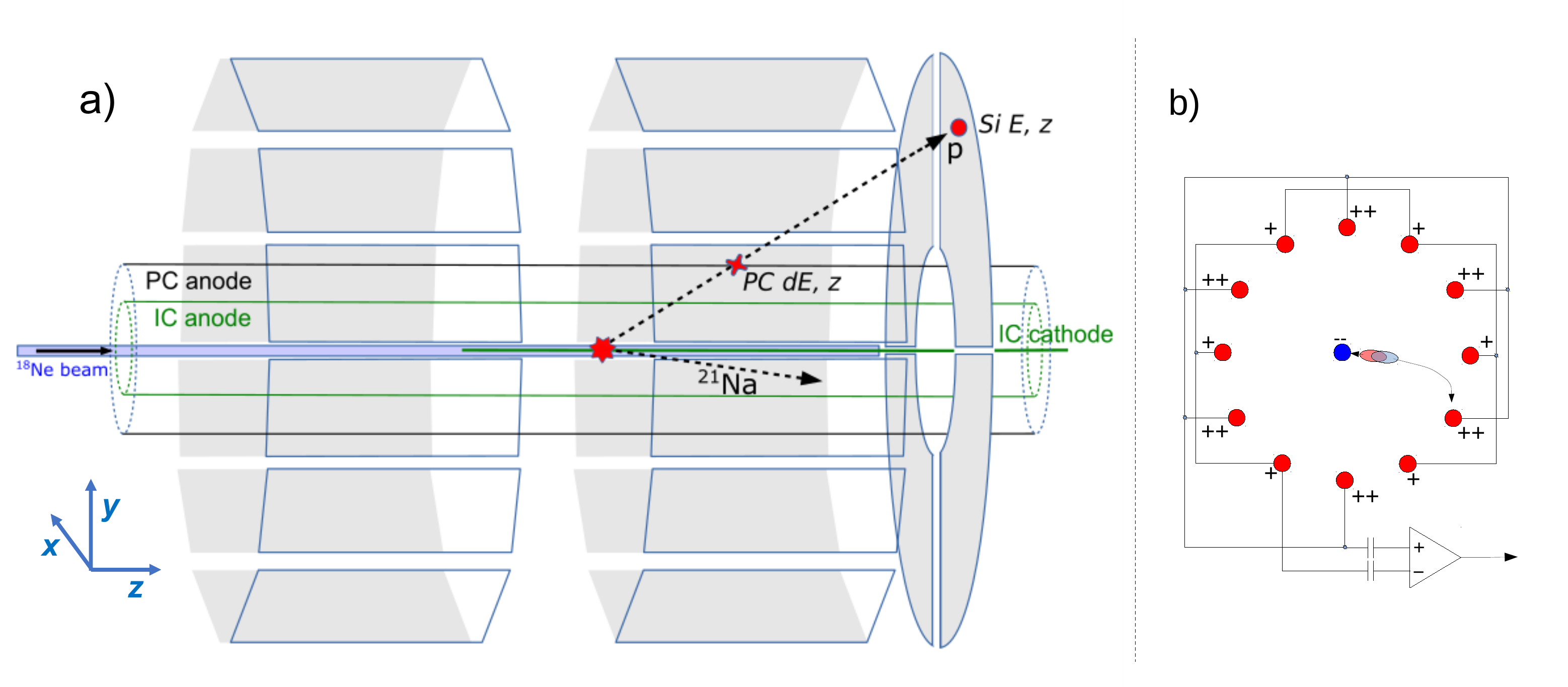}
	\end{minipage}
	\caption{Panel (a): Schematic representation of the ANASEN detector system. The $(\alpha,p)$ proton tracks are reconstructed from the signals of a resistive proportional-counter anode in coincidence with their residual energy and position on a multi-strip silicon detector. The cylindrical ion-chamber component measures the ionization induced by the beam particles and heavy ion reaction residues (see text). Panel (b): Schematic of the inner ion-chamber component of ANASEN, 
	which uses a cylindrical field between the central Cathode and a grid of 48 Anode wires (only 12 shown here for clarity), which are biased at positive potentials of alternating magnitude and read out in a differential scheme. } 
	\label{fig:anasen_schematic}
\end{figure*}

The $^{18}$Ne($\alpha,p$)$^{21}$Na reaction was studied in inverse kinematics using a beam of the short-lived isotope $^{18}$Ne, produced with the RESOLUT radioactive beam facility \cite{resolut} at the John D. Fox Superconducting Accelerator Laboratory of Florida State University. The $^{18}$Ne beam was produced via the $\mathrm{^{16}O(^3He,n)^{18}Ne}$ reaction and selected at 80~MeV and at an average beam intensity of 2$\cdot$10$^3$ pps, which constitutes 14\% of the particles delivered. Offline, the events associated with the $^{18}$Ne beam particles were identified and discriminated from the $^{16}$O contaminant to 98\% purity through individual time-of-flight signals relative to the accelerator RF-reference signal.  

The experiment was performed with the active-target detector Array for Nuclear Astrophysics and Structure with Exotic Nuclei (ANASEN). A schematic of the detection scheme is displayed in Fig.\ \ref{fig:anasen_schematic}, and more details on the design and operation of ANASEN are described in Ref.\ \cite{anasen-nim}. ANASEN was filled with $^{4}$He as the target gas including a 4\% admixture of CO$_{2}$ as a quenching gas, at a pressure of 379 Torr. The beam particles entered the gas-volume through an 8.9~$\mu$m Kapton window and continuously lost energy in the target gas, such that events from the energy range of interest occurred in an area of high geometric efficiency. The ANASEN detector systems were arranged in concentric layers surrounding the beam axis in a cylindrical geometry, starting from an inner set of 24 position-sensitive proportional-counter (PC) anodes at 3.75~cm radius, to a layer of 24 position-sensitive silicon-strip detectors of the Micron Semiconductor ``Super-X3'' design at a radius of 8.9~cm. In addition, 4 silicon-strip detectors of the Micron-Semiconductor ``QQQ3'' design cover forward laboratory angles in an annular geometry. The $(\alpha,p)$ protons were first detected in the proportional counter as they left the central beam volume, using the charge division on the resistive anode wire to determine the z-coordinate of interaction on the beam axis, along with the characteristic energy loss of the particles. The protons proceeded to the outer perimeter of ANASEN, where they were stopped in the silicon detectors, which allowed to determine the residual energy and position.  

For this experiment a new component was added to ANASEN, a high-rate, compact ion chamber, sensitive to heavy ions traveling within 2.2~cm of the beam axis, including both beam particles and reaction residues. The ion chamber drift field is created between a negatively biased (-200 V) tungsten-wire cathode on the beam axis and 48 anode wires surrounding it at 2.2~cm cylindrical radius, as schematically represented in Panel (b) of Fig.\ \ref{fig:anasen_schematic}. A differential read-out scheme is applied between odd- and even-numbered anode wires, which were biased at 150~V and 250~V, respectively. The drifting electrons induce differential signals only once they approach the anode radius, which creates the effect of a Frisch-grid and allows the ion chamber to operate up to 10$^5$ counts/second.

Because of the geometry of the thin tungsten wire, the ion chamber is referred to as the Needle-IC. The tungsten-wire cathode is mounted on a retractable shaft inserted from the end of the detector chamber, such that the active volume of the ion chamber is adjustable. During the experiment, the needle was inserted to a depth of 18~cm into the active ANASEN volume. It should be noted that the 0.5-mm-diameter tungsten-wire is extending into the path of the beam, which had a diameter of $\approx$ 1.5~cm. While this aspect has the potential to create additional background scattering and reaction events, those were suppressed by requiring the expected recoil-energy correlation in the data analysis.     

The variable insertion depth of the Needle was also used to determine the range of beam particles in the target gas independently. The result of this range measurement was compared to calculations with the program SRIM \cite{Ziegler10} and LISE$++$ \cite{lise}. The energy-loss measurement was consistent with LISE$++$, but the SRIM energy-loss rates had to be corrected by a factor of $\approx$ 2. The resulting energy-loss profile was found to be consistent with the profile determined from the kinematics of two-proton coincidence events described in section~\ref{sec:level2-2pevents}. 

The energy calibrations of the silicon detectors were determined using $\alpha$ particles from a $^{228}$Th source in vacuum, as well as by scattering accelerator-provided $\alpha$ and proton beams off a thin tantalum foil mounted on-axis within the detector gas. The position of the scattering target was also varied along the beam axis, in order to calibrate the vertex-position reconstruction from the proportional-counter anodes in conjunction with the silicon detectors. Based on these calibration runs, the reaction vertex could be reconstructed with a resolution of about 2.5~cm FWHM, limited by the position resolution of the proportional counter and the beam dispersion.

\section{\label{sec:level1-Analysis}Data Analysis}
The experiment was designed to measure the excitation function of the $^{18}$Ne($\alpha,p$)$^{21}$Na reaction reaching to the low energies relevant for nuclear astrophysics. The energy-level scheme of the reactants, displayed in Fig.\ \ref{reacscheme}, shows that, owing to the low proton threshold in $^{22}$Mg, a range of excited states in $^{21}$Na can be populated, including states above the proton-decay threshold of $^{21}$Na. 

\begin{figure}[ht]
	\centering
	\begin{minipage}{0.49\textwidth}
		\centering
		\includegraphics[width=\textwidth]{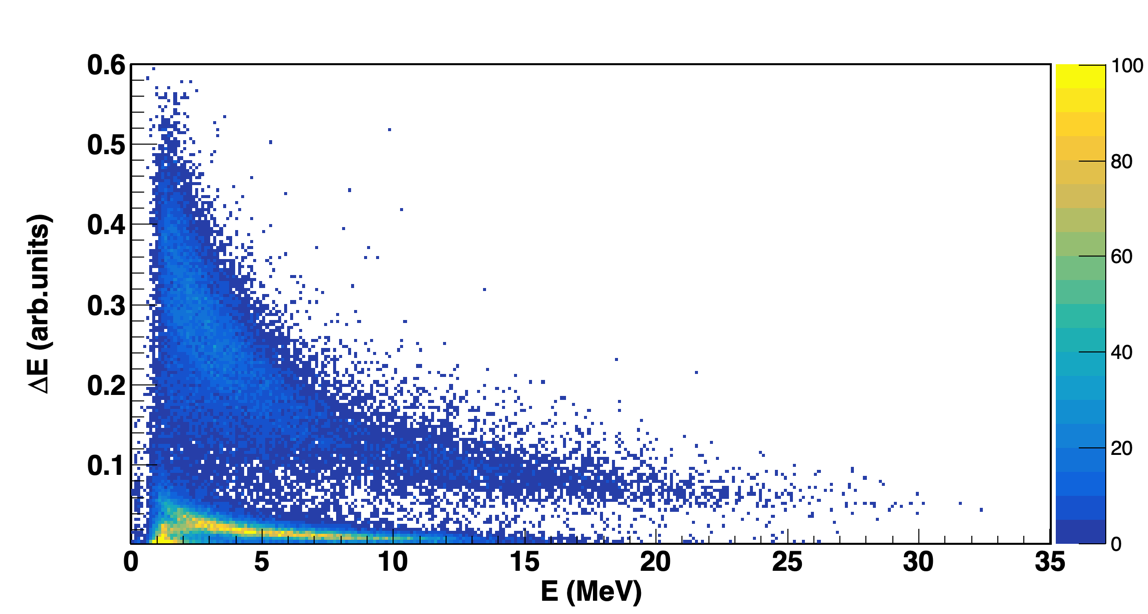}
	\end{minipage}
	\begin{minipage}{0.49\textwidth}
		\centering
		\includegraphics[width=\textwidth]{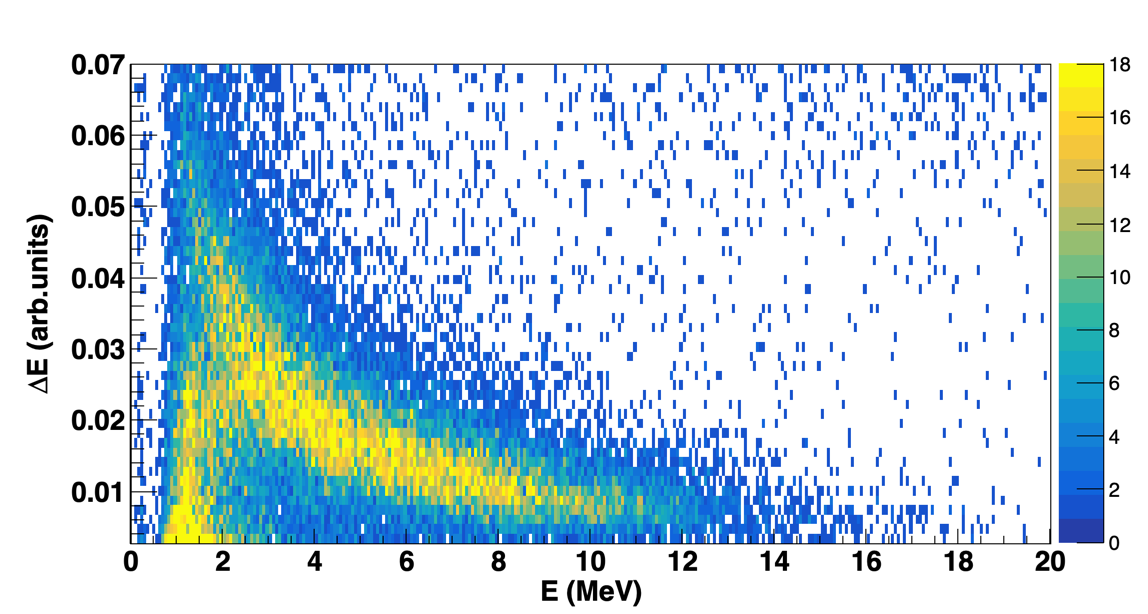}
	\end{minipage}
	\caption{Top: Energy-loss signals of the proportional-counter anodes vs.\ the coincident silicon detector signal. The energy-loss information ($\Delta E$) is derived from the PC signals, corrected for the angle-dependent thickness of the active detector volume, and the residual energy (E) is measured in the Si detectors. Protons and $\alpha$--particles show distinct, separate correlation groups. Bottom: Display of the same (Top) histogram expanded to show the region of proton signals.}
	\label{ede}
\end{figure}

\begin{figure}[htp]
	\centering
	\includegraphics[width=0.5\textwidth]{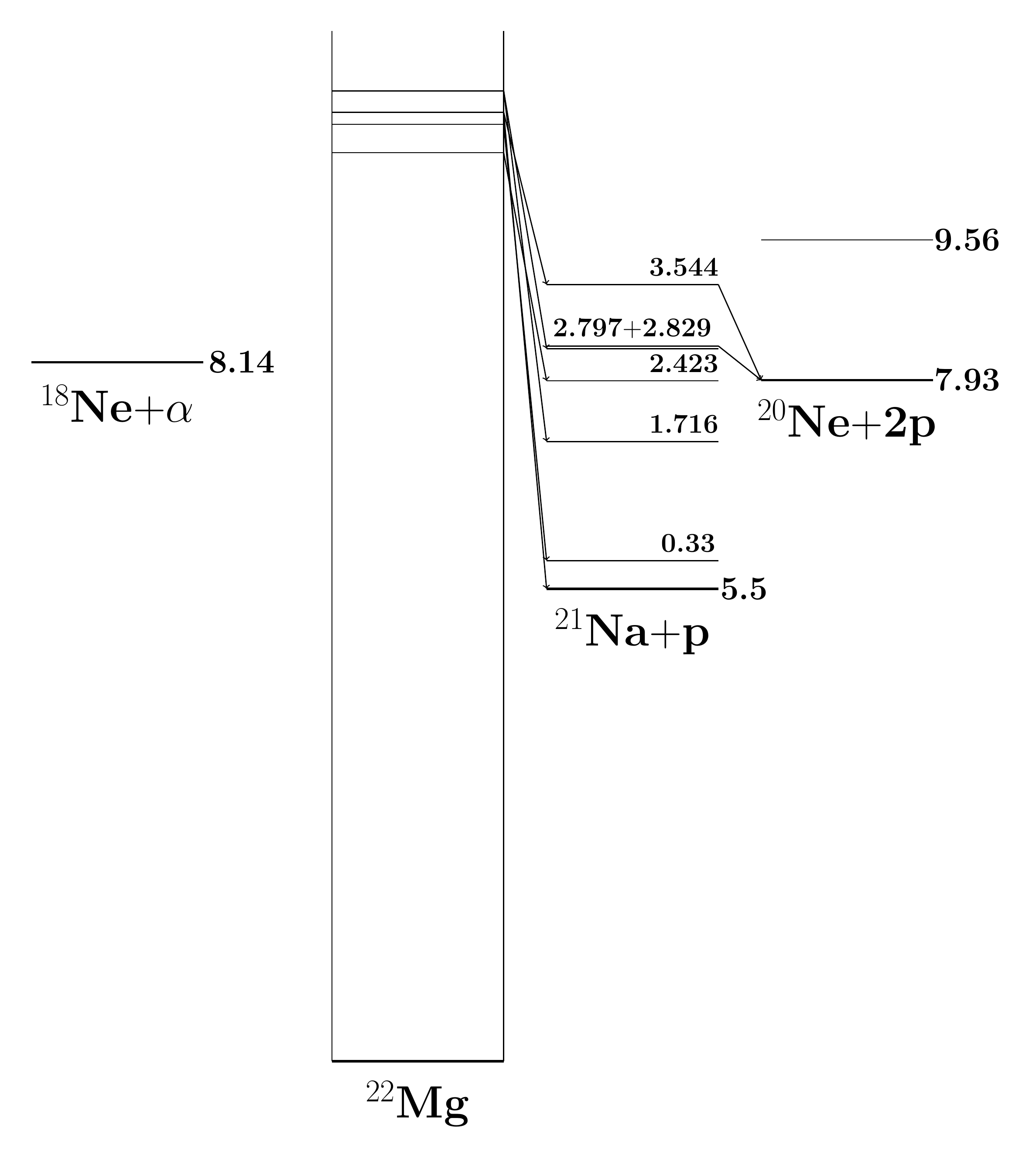}
	\caption{Level schemes of $^{22}$Mg, $^{21}$Na and $^{20}$Ne including the respective reaction and decay energy thresholds in MeV. } 
	\label{reacscheme}
\end{figure}

\subsection{\label{sec:level2-sp} Single-Proton Events}
The continuous measurement of an excitation function in the ANASEN detector requires the event-by-event determination of the reaction energy, which is obtained by extrapolation of the proton trajectory to the reaction vertex on the beam axis, and the calculation of the $^{18}$Ne beam-particle energy loss up to that point. We refer to this reconstruction method as $E_{track}$. The identity of the detected protons was established through their characteristic energy-loss signals, as represented in the $\Delta E$-$E$ particle identification plot in Fig.\ \ref{ede}. The proton energies were derived from the residual energy measured in the silicon detectors by correcting for the energy loss in the path from the reaction vertex to the silicon detector.

\begin{figure}[htp]
	\centering
	\includegraphics[width=0.5\textwidth,height=5cm]{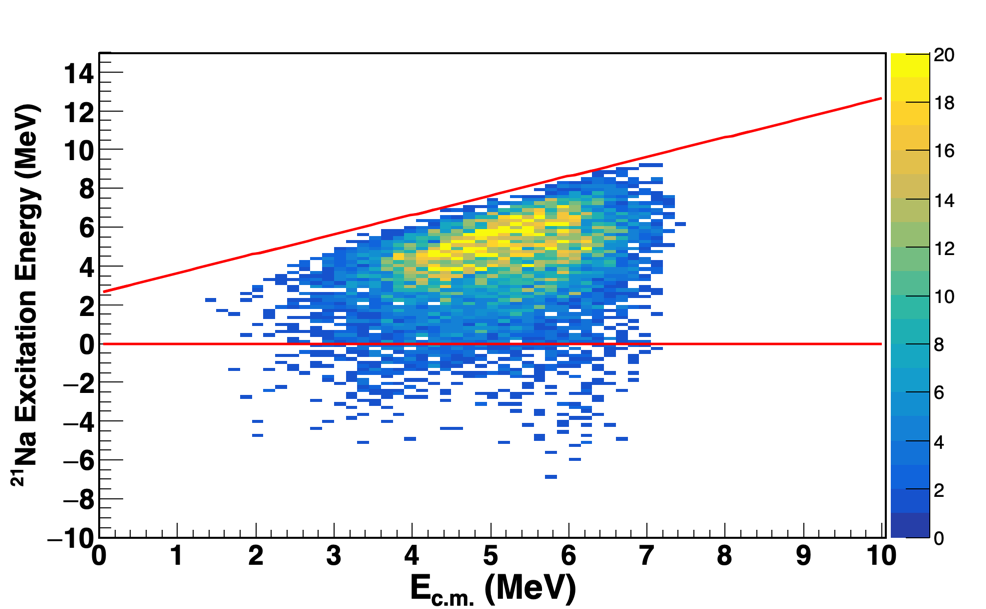}
	\caption{Single-proton events: Distibution of $^{21}$Na excitation energies with respect to the reaction energy in the center-of-mass frame, reconstructed using the E$_\mathrm{track}$ method. The two lines delimit the kinematically allowed event range.} 
	\label{fig:21exvsbeam}
\end{figure}

The single-proton events, selected through the $\Delta E$-$E$, and Needle-IC gates were analyzed using the $E_{track}$ method to determine the reaction energy. The gates obtained from the Needle-IC will be described in Section \ref{sec:level2-needle}. The angle and the energy of the protons were also used to reconstruct the associated $^{21}$Na excitation energy. The resulting distribution of reaction and excitation energies is presented in Fig.\ \ref{fig:21exvsbeam}. 
A wide distribution of reaction energies is observed, down to about 2~MeV in the center of mass frame. The distribution also shows that lower $^{21}$Na excitation energies are populated in correlation with lower reaction energies, constrained by the $E_x \leqslant E_{c.m.} + Q_{value}$ limit. Also visible are some events which were reconstructed at negative excitation energies. These events can be likely attributed to fusion-evaporation reactions on the CO$_2$ component of the target gas, or as an effect of the poor resolution (1.45~MeV FWHM) of the tracking reconstruction. Because of the limited resolution, we are not able to separate each individual state in $^{21}$Na, but identify three regions, the ground-state region including the ground state and first-excited state at 0.33~MeV, the bound-state and the unbound-state regions.

\subsection{\label{sec:level2-2pevents}Two-Proton Events}
\begin{figure}[htp]
	\centering
	\includegraphics[width=\columnwidth]{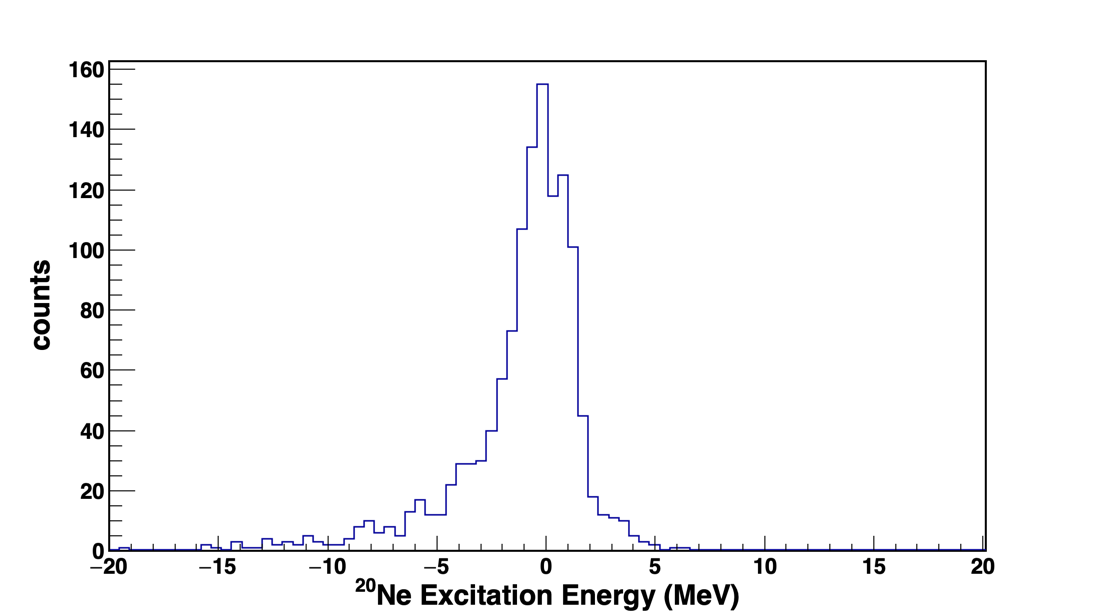}
	\caption{Two-proton events: $^{20}$Ne excitation-energy distribution, using the E$_\mathrm{track}$-reconstruction of the reaction energies. } 
	\label{fig:20neqvalue}
\end{figure}

The experiment detected a significant number of coincident two-proton events, which originate from the population of proton-unbound states in $^{21}$Na. These data were analyzed using the $E_{track}$ reconstruction of the $^{18}$Ne energy, and the proton energies and angles to reconstruct the $^{20}$Ne excitation energy. The resulting spectrum is displayed in Fig.\ \ref{fig:20neqvalue}. The data is dominated by a peak at the ground-state energy of $^{20}$Ne. This peak position is very sensitive to the energy-loss calibration and represents an independent confirmation of its validity.  

The two-proton events also allow to reconstruct the excitation energies of $^{21}$Na and $^{22}$Mg relative to the $^{20}$Ne ground state, since the $^{18}$Ne($\alpha,2p$)$^{20}$Ne reaction proceeds dominantly through the $^{20}$Ne ground state and thus the $Q$-value is known. This reconstruction method, called $E_{kine}$, achieves better resolution than the $E_{track}$ method, around 500~keV FWHM, as it reduces the reliance on the z-coordinate of reaction vertex, whose reconstruction is hampered by the $\approx$ 1.5 cm beam diameter.

\begin{figure}[htp]
	\centering
	\includegraphics[width=\columnwidth]{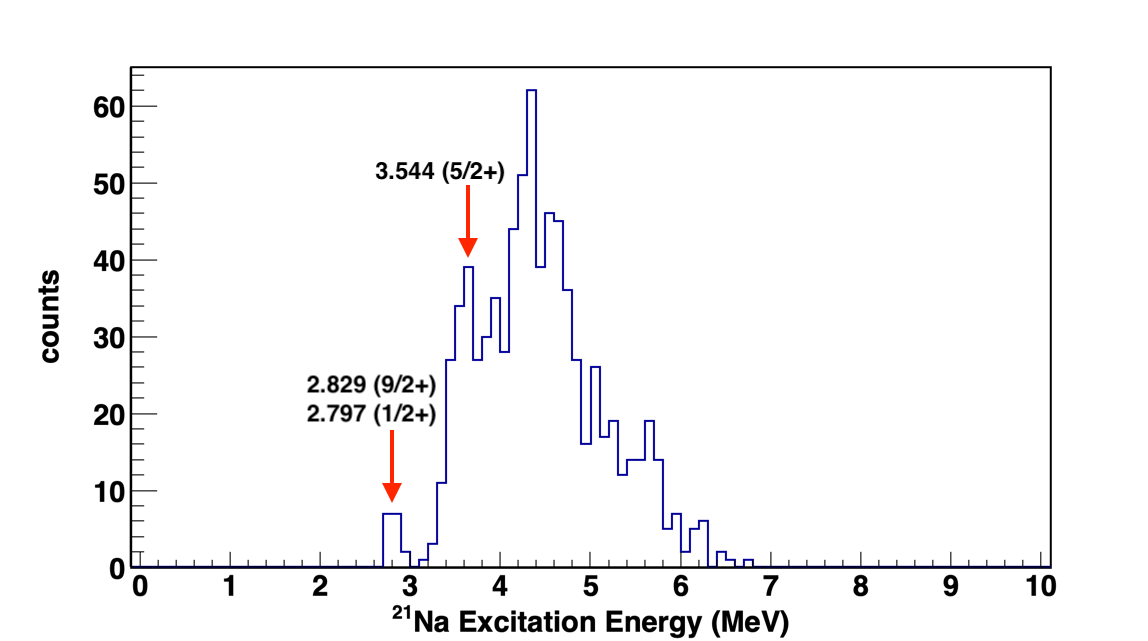}
	\caption{Two-proton events: $^{21}$Na excitation-energy distribution reconstructed from the $^{20}$Ne+p energies and the E$_\mathrm{kine}$ method. } 
	\label{fig:21naex_2p}
\end{figure}

\begin{figure}[h]
\includegraphics[width=\columnwidth,height=4.8cm]{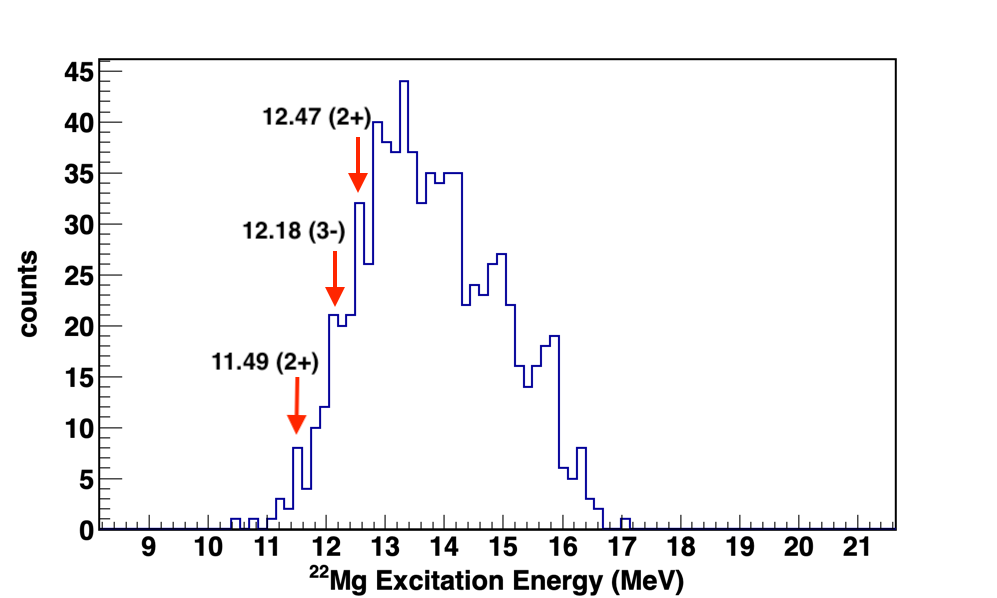}
\caption{Two-proton events: $^{22}$Mg excitation-energy distribution reconstructed from the $^{20}$Ne+2p energies and the E$_\mathrm{kine}$ method.}
\label{fig:22mgex_2p}
\end{figure}

In Fig.\ \ref{fig:21naex_2p} the spectrum of the $E_{kine}$-reconstructed excitation energies of $^{21}$Na is shown, with the events gated on the ground-state peak of $^{20}$Ne. The doublet of states at 2.8~MeV and the singlet 3.54~MeV peak can be identified right above the 2$p$-threshold, while a higher density of states is observed beyond this point. In Fig.\ \ref{fig:22mgex_2p} the $E_{kine}$-reconstructed $^{22}$Mg excitation-energy spectrum is displayed, again gated on the $^{20}$Ne ground-state peak. While it is difficult to assign individual resonances to the observed structures, the lower part of the spectrum seems to be consistent with the known natural-parity resonances in $^{22}$Mg. 

\subsection{\label{sec:level2-needle}Reaction-Residue Analysis with Needle-IC}

The Needle-IC signals were used to suppress background events from scattered particles or fusion-evaporation reactions on the CO$_2$ admixture in the target gas, while accepting the $^{18}$Ne($\alpha,p$)$^{21}$Na events of interest. The signal interpretation, however,  is complicated by the fact that it includes the ionization by the beam-particle up to the reaction point, in addition to the ionization created by the reaction residue. Since the observed reactions have a $Q$-value close to zero, and some energy is carried away by the emitted protons, the reaction-residue energy is necessarily lower than the beam-particle energy at the reaction point. In effect, the Needle-IC signals will be maximal when the reaction occurs at the minimal reaction energies, where the signals are dominated by the beam-particle ionization.

\begin{figure}[htp]
	\centering
	\includegraphics[width=\columnwidth]{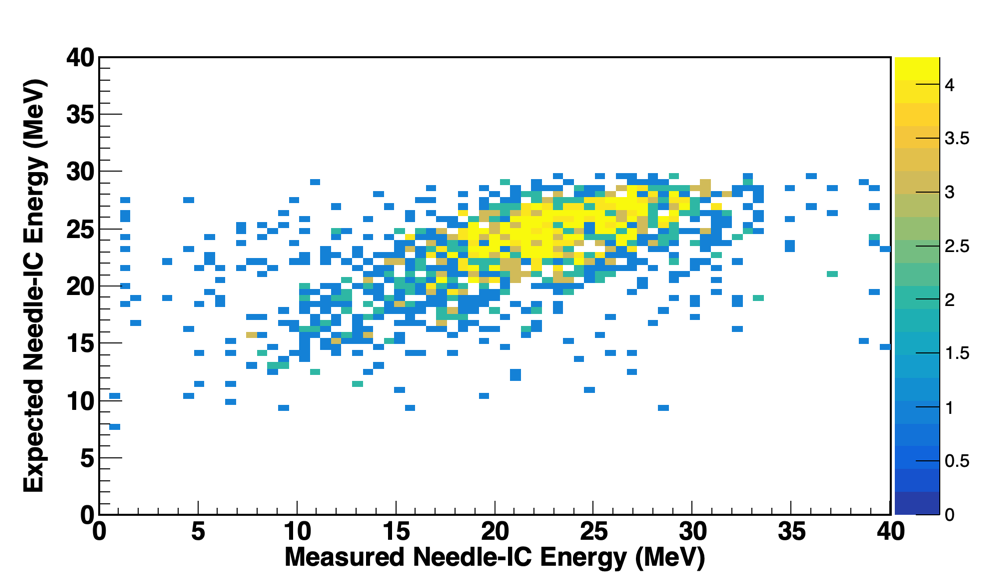}
	\caption{Two-proton events: Correlation between ``expected'' and ``measured'' Needle-IC signals, gated on the $^{20}$Ne ground-state peak.} 
	\label{needle2p}
\end{figure}

We analyze the Needle-IC signals by calculating the expected detector response and comparing it with the measured signal. The expected signal is determined from the initial energy of the reaction residue, based on the event kinematics, plus the integrated energy loss of the beam up to the point of reaction. Both components are analyzed for the ionization deposited within the ion chamber’s active volume. The correlation between the ``expected'' and the ``measured'' Needle-IC energy signals for the two-proton events is displayed in Fig.\ \ref{needle2p}. The events close to the diagonal are the ($\alpha,2p$) events of interest, while the events scattered away from the main group, not showing the expected recoil-energy correlation, correspond to background. 

\begin{figure}[h]
	\centering
	\includegraphics[width=0.49\textwidth]{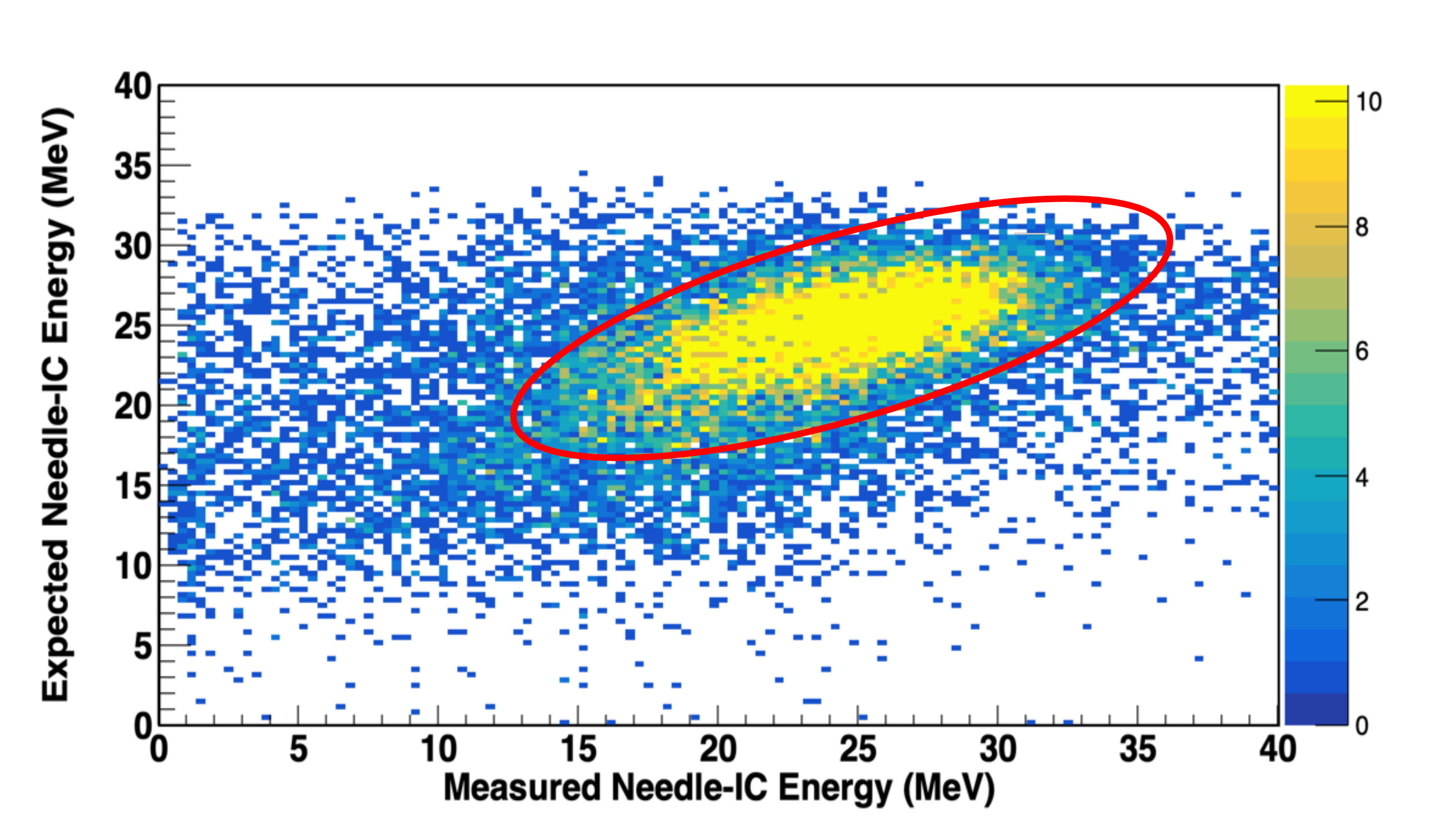}
	\caption{Correlation between ``expected'' and ``measured'' Needle-IC signals, for the single-proton events. The correlated events in the main diagonal, inside the gate, were chosen for further analysis. } 
	\label{fig:needle}
\end{figure}

In order to determine the efficiency of the Needle-IC detector and the recoil-correlation analysis, we examined how many of the 2p-events leading to the $^{20}$Ne ground-state peak also show the expected Needle-IC signals in coincidence. The efficiency of recoil detection was determined as the ratio of both numbers and a smooth function of the reaction energy was extracted from this data as a model for the ion-chamber efficiency. The Needle-IC gate suppresses a significant number of events at higher energies, but leaves the energy region of interest unaffected, consistent with 100\% efficiency below $E_{c.m.}\approx$~4.5~MeV. The same Needle-IC analysis was also applied to the single-proton events. In Fig.\ \ref{fig:needle} the ``expected'' signal is plotted against the ``measured'' Needle-IC energy signal for single-proton events. Again, we observe a group of correlated events on the main diagonal, with scattered background surrounding it. The figure also illustrates the region of events chosen for analysis.

\subsection{\label{sec:eff}Determination of Detector Efficiency}
The detector efficiency was calculated through a Monte-Carlo simulation of the event kinematics and the resulting detector response. The reaction energy was randomly chosen within the range of interest for each simulated $^{18}$Ne($\alpha,p$)$^{21}$Na event. The event vertex location associated with that energy was determined by the same energy-loss tables used in the experimental data analysis. This vertex position was offset by random Gaussian distributions in x, y and z to reproduce the distribution of the $\approx 1$-cm-diameter-beam, as well as the straggling accumulated through the energy loss at the vertex position. At the lowest reaction energies, the accumulated straggling effects result in $\sim$1.5~cm diameter for the final beam spot size. The $^{21}$Na excitation energy was chosen randomly within the kinematic limits, as well as the proton-decay angle, chosen to be isotropic in the center-of-mass system. In addition, if the $^{21}$Na excitation energy was falling above the 2$p$-threshold at 2.43~MeV, an additional proton was generated in the same way. One or both randomly chosen protons were projected from their generation point through the gas volume towards the silicon detectors. The proton was accepted in the simulation if an active area of the silicon detectors was reached, with energy greater than the detection threshold of $\approx$ 0.5 MeV after taking the energy losses into account.   

\begin{figure}[htb]
	\centering
\includegraphics[width=0.5\textwidth]{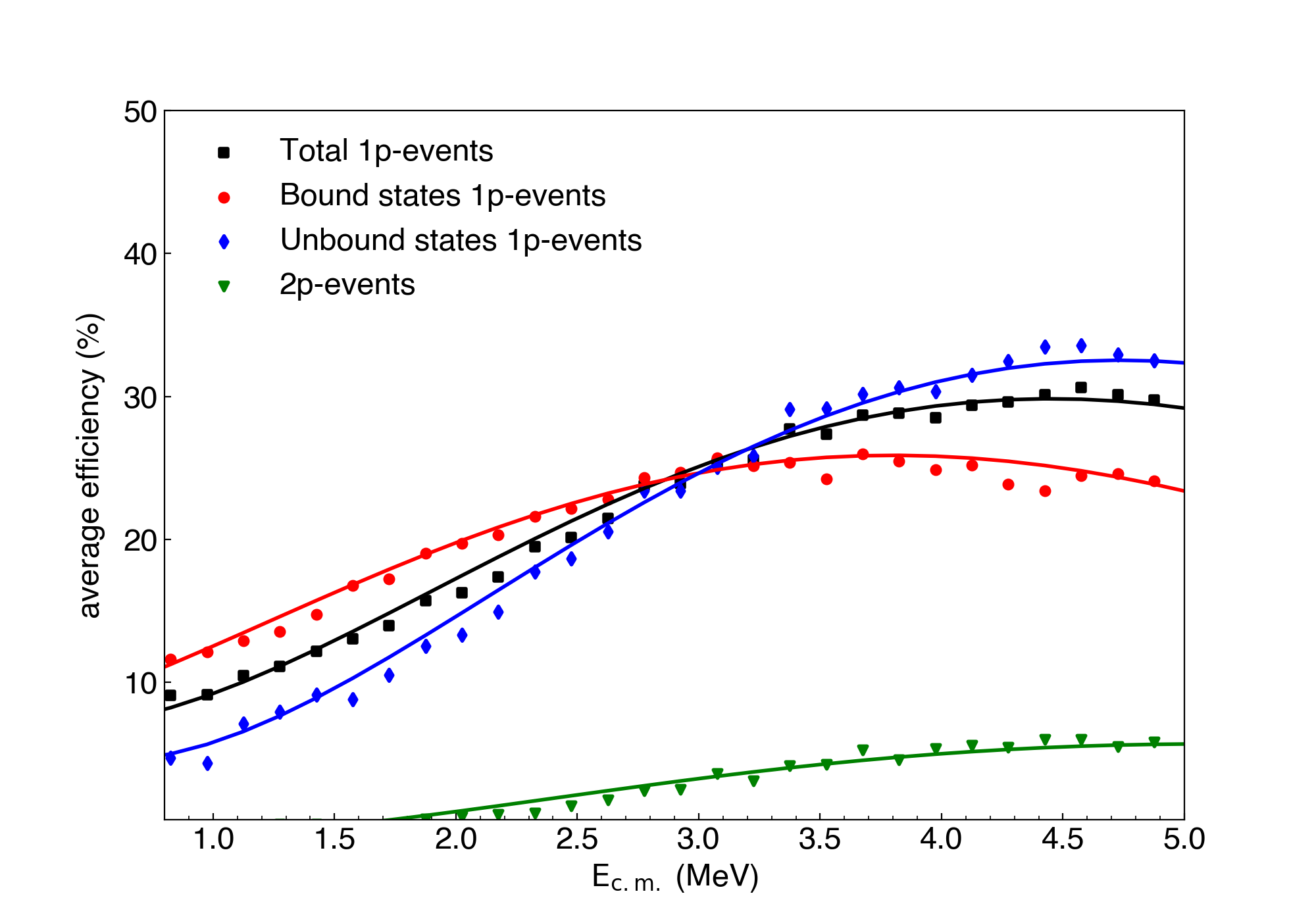}
	\caption{Average detection efficiencies from a Monte-Carlo simulation of the ($\alpha$,$p$) reaction and the corresponding detector response, averaged over various $^{21}$Na excitation-energy regions. Also shown is  
	the efficiency for the coincident detection of two protons from the ($\alpha$,2$p$) reaction}  
	\label{fig:efficiency}
\end{figure}

In Fig.~\ref{fig:efficiency}, the simulated ($\alpha$,$p$) detection efficiency, averaged over all allowed $^{21}$Na excitation energies, is presented, as well as the efficiency for the coincident detection of two protons. All efficiency curves show a slowly varying dependence on the reaction energy. The average detection efficiency peaks at $\approx$ 30$\%$, for events at $E_{c.m.}=$ 5~MeV and drops continuously to $\approx$ 12$\%$ for the lowest-energy events. The efficiency for the coincident detection of two protons peaks at $\approx$ 5$\%$ and drops to $\approx$ 1$\%$ for the lowest-energy region. Fig.~\ref{fig:efficiency} also shows the average single-proton efficiency for bound-state and unbound-state regions. The unbound-state events are detected with a significantly higher efficiency, as both protons have similar chances of being detected. Since there is no experimental signature to determine which of the protons was detected, a fraction of the resulting single-proton events will be incorrectly assigned a too-high $^{21}$Na excitation energy, which was taken into account in the efficiency-simulation. This effect however diminishes at lower reaction energies, and has minimal impact on the analysis of the overall excitation function for $E_{c.m.}<$~4~MeV. 

\subsection{\label{sec:level2-gaussian} Gaussian Unfolding of Spectra}

\begin{figure}[htp]
	\centering
\includegraphics[width=0.5\textwidth]{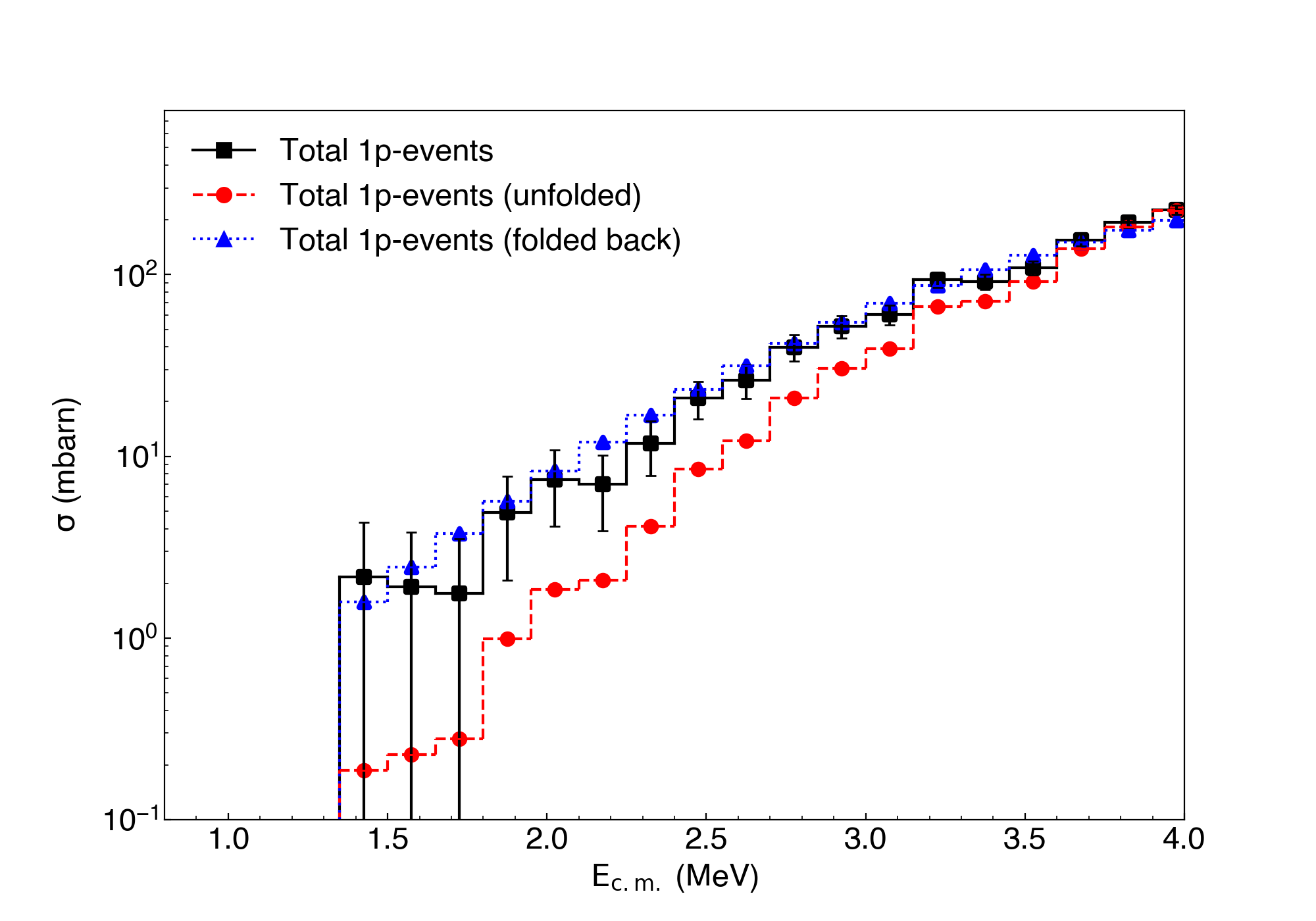}
	\caption{Demonstration of the unfolding of the observed experimental cross section spectrum with respect to its Gaussian resolution, following the 2-step iteration of the Richardson-Lucy algorithm. The process is verified by folding back the unfolded spectrum with the Gaussian resolution. The error bars on the unfolded and folded-back spectrum are omitted for clarity.} 
	\label{xsections_unfolded}
\end{figure}

The reaction-energy resolution of the experiment from the $E_{track}$ method, $1.45\pm 0.07$ MeV FWHM in c.m., was determined by the Monte-Carlo simulation, where the simulation-generated reaction energy is compared to the $E_{track}$-energy reconstructed for the same simulated event. This limited energy resolution also leads to background, through the folding of the higher-energy signals, detected with larger cross sections, into the lower-cross section region of interest. To quantify and correct for this effect, we unfold the observed experimental spectra through a 2-step iteration of the Richardson-Lucy algorithm \cite{Zech13}, assuming a Gaussian response function with $1.45$ MeV FWHM.
In Fig.\ \ref{xsections_unfolded} the raw experimental cross section for single-proton events is shown and compared to the resulting, unfolded spectrum. We verify that the inverse process, folding the unfolded spectra with a Gaussian, leads to the observed spectrum within the uncertainty limits, as shown in Fig.\ \ref{xsections_unfolded}. Additionally, the 
systematic uncertainty of the unfolding procedure owing to the 0.07~keV uncertainty of the width parameter 
is much smaller than the statistical errors.
In effect, the unfolding procedure is lowering the extracted cross sections for the lowest energies while leaving the higher-energy spectra mostly unchanged. In the remainder of the paper, we focus on the unfolded distributions. 

\section{\label{sec:level1-Results}Results}

\begin{figure}[ht]
	\centering
	\begin{minipage}{0.49\textwidth}
		\centering
	\end{minipage}
	\begin{minipage}{0.49\textwidth}
		\centering
\includegraphics[width=\textwidth]{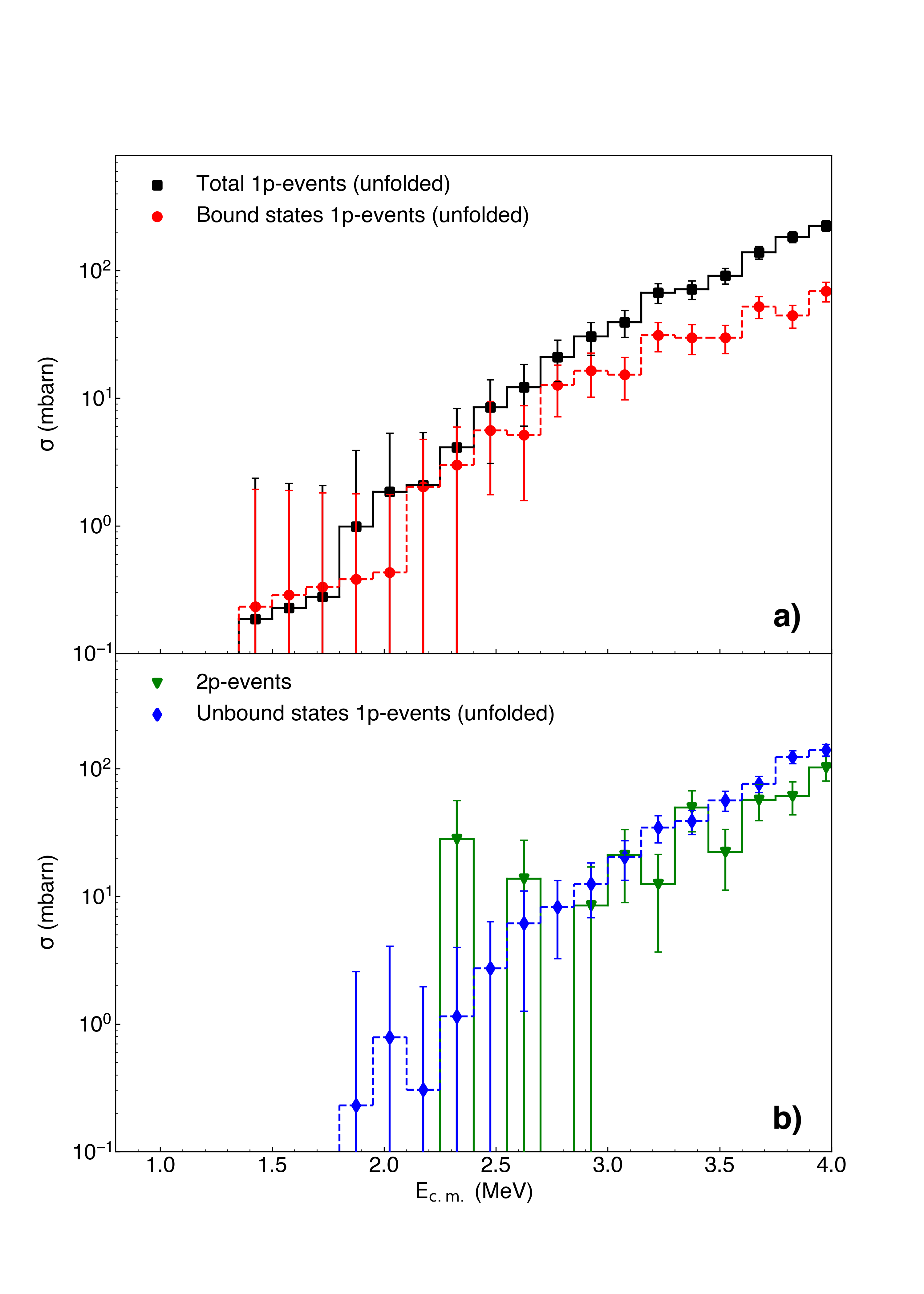}
	\end{minipage}
	\caption{Panel~(a): Cross sections from single-proton events determined from this work and corrected for the experimental energy resolution. The total cross section (black) is compared to cross section for bound states only (red). Panel~(b): The unbound states cross sections from the single-proton events analysis, corrected for the experimental energy resolution,  compared to the cross section obtained from the independent analysis of the coincident 2$p$-events. Due to higher resolution, the 2$p$-events cross section is not subjected to the Gaussian unfolding procedure. }
	\label{xsections}
\end{figure}

The cross sections were determined from the number of detected events for a given reaction-energy interval, the simulated efficiency, the total number of beam particles, and the effective areal target thickness for the corresponding energy interval. The total number of beam particles was determined from the number of particles detected in the Needle-IC detector, whose events were sampled independently, once for every thousand triggers. The areal target density was calculated from the target-gas pressure, the target composition, and the beam-path length associated with the individual reaction-energy interval, consistent with the energy-depth profile used in the $E_{track}$ reconstruction. The events were separated into intervals of $^{21}$Na excitation energy, beginning with the ground-state region between -1 and +1 MeV, over the bound-state region up to 3 MeV, to the higher excitations, which were separated into intervals of 0.4 MeV width, up to excitation energies of 7 MeV. These data-intervals were corrected for the average efficiency over the equivalent interval of the simulated events. The efficiency-corrected data was then combined to the $^{21}$Na excitation regions discussed below and 
subjected to the Gaussian unfolding procedure. 

Panel (a) of Fig.\ \ref{xsections} displays the $^{18}$Ne($\alpha,p$)$^{21}$Na cross sections determined in this work for all single-proton events, separating the cross sections for population of the bound-state region. As expected, the cross section data shows a steep drop towards lower energies, reaching the sensitivity limit of the experiment at around 2.5~MeV (c.m.). At the lowest energies, the total cross section is dominated by population of the $^{21}$Na bound states but shifts more to unbound states above 3~MeV (c.m.). Panel (b) of Fig.\ \ref{xsections} shows the cross section of the two-proton events which are detected in coincidence and corrected for their corresponding efficiency shown in Fig.\ \ref{fig:efficiency}. These events were analyzed independently as discussed in section \ref{sec:level2-2pevents} representing the population of unbound excitations. While the statistical uncertainties of the coincident two-proton data are much larger, this data has a higher resolution of $\approx 500$~keV due to the $E_{kine}$ method, and is therefore not subjected to the Gaussian-unfolding procedure. This data is compared to the cross section extracted from single-proton events, associated with the $^{21}$Na unbound-excitation energies. The cross-section of the single-proton events is slightly higher than the cross section extracted from 2p-events, but nearly consistent within the error bars. 

\section{\label{sec:discussion}Discussion}

\begin{figure}[h]
	\centering
	\begin{minipage}{0.49\textwidth}
		\centering
	\end{minipage}
	\begin{minipage}{0.49\textwidth}
		\centering
		\includegraphics[width=\textwidth]{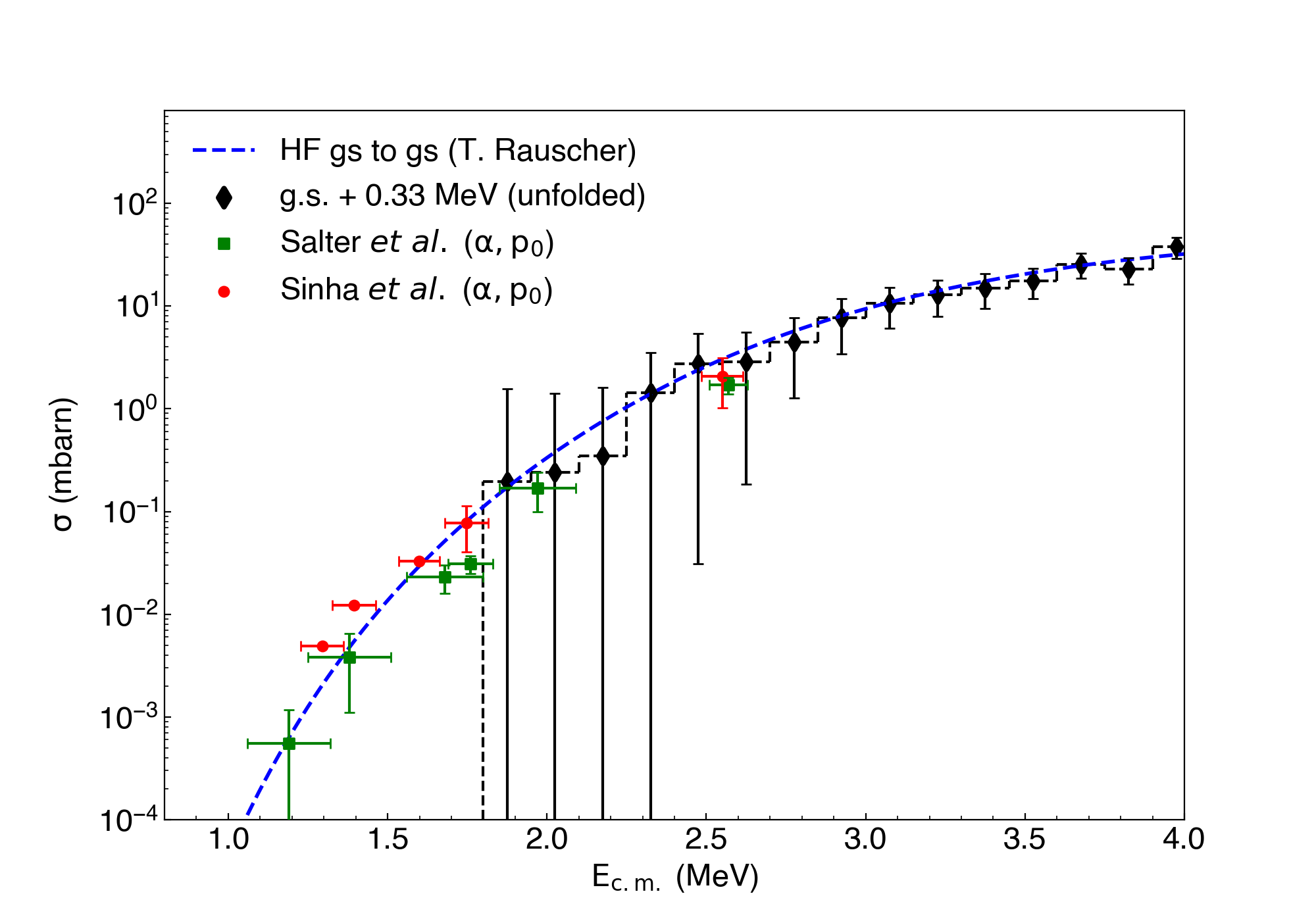}
	\end{minipage}	
	\caption{The ground state cross section, including the  first excited state at 0.33~MeV, after resolution adjustment. Previously collected data by Sihna {\it et al.} \cite{ANL} and Salter {\it et al.} \cite{salter} that correspond to the ground state transition, and the Hauser-Feshbach calculation for ground-state to ground-state transitions are plotted for comparison.}
	\label{fig:xsec_gs_comparison}
\end{figure}

Experimental data on the $^{18}$Ne($\alpha,p$)$^{21}$Na reaction were obtained by Salter {\it et al.} \cite{salter} in a measurement of the time-inverse reaction $^{21}$Na($p,\alpha)^{18}$Ne and applying the principle of detailed balance. This interpretation, however, can only be applied to reactions leading to the $^{21}$Na ground state, while even at the lowest reaction energies, excited-state population is energetically allowed and may contribute substantially to the reaction rate. Our experiment allows to select events associated with the unresolved $^{21}$Na ground state and the 0.33~MeV first-excited state.  
Fig.\ \ref{fig:xsec_gs_comparison} shows the thus-obtained cross sections and compares them to the corresponding cross sections extracted from Salter {\it et al.} \cite{salter} and Sihna {\it et al.} \cite{ANL}. Within the shared energy interval and uncertainties, both experiments agree. Due to the unresolved first-excited-state contributions, we expect that our cross section is slightly larger when compared to the time-reverse measurements. Also shown is a Hauser-Feshbach calculation of the ground-state to ground-state cross section, performed with the program SMARAGD \cite{smaragd1, smaragd2}. The Hauser-Feshbach calculation shows the same energy dependence as our data and the data from the time-reverse reaction experiments \cite{ANL, salter}. 

\begin{figure}[ht]
	\centering
	\begin{minipage}{0.49\textwidth}
		\centering
	\end{minipage}
	\begin{minipage}{0.49\textwidth}
		\centering
\includegraphics[width=\textwidth]{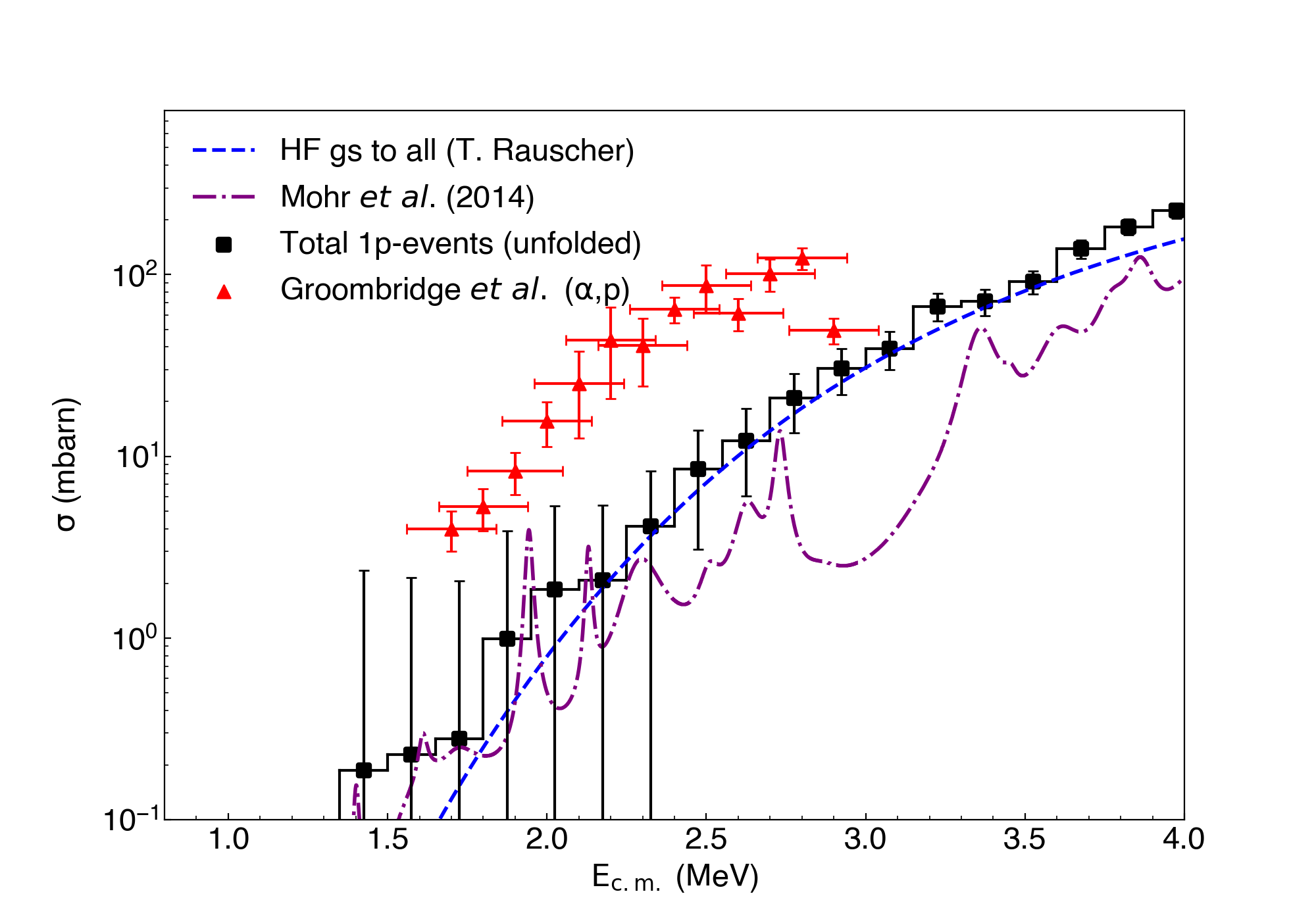}
	\end{minipage}
	\caption{Comparison of the total ($\alpha$,$p$) cross section of this work with the previously obtained ($\alpha$,$p$) cross section by Groombridge {\it et al.} \cite{groom}, as well as with theoretical calculations by Mohr {\it et al.} \cite{mohr2014} and the Hauser-Feshbach code SMARAGD \protect\cite{smaragd1, smaragd2}.}
	\label{fig:xsec_comp_results}
\end{figure}

As discussed in the introduction, the $^{18}$Ne($\alpha,p$)$^{21}$Na cross section has been studied before by Bradfield-Smith {\it et al.} \cite{bradfield} and Groombridge {\it et al.} \cite{groom}. There, the experimental results were summarized as a table of resonance parameters and widths. Fig.\ \ref{fig:xsec_comp_results} displays the cross section derived from the resonance parameters of Ref.\ \cite{groom}, in comparison to our result. Our experiment  shows a much lower cross section, by close to an order of magnitude.

Also displayed in Fig.\ \ref{fig:xsec_comp_results} is the cross section derived by Mohr {\it et al.} \cite{mohr2014} from a survey of $^{22}$Mg resonance parameters, obtained in indirect experiments \cite{Matic09, Chae09, jjhe2013, zhang2014}, or calculated assuming mirror-symmetry in the wave functions of corresponding $^{22}$Mg and $^{22}$Ne levels. Where no spectroscopic information was available for the $^{22}$Mg levels, these were calculated from a Porter-Thomas distribution following Ref.\ \cite{porter}. While the resulting cross-section curve by Mohr {\it{et al.}} \cite{mohr2014} is consistent with our data in the lower-energy range, the curve is systematically too low at higher energies. This systematic deviation can be attributed to underestimating the contributions to higher-lying $^{21}$Na states, which evidently dominate our experimental data at higher energies, or to contributions of unknown or mis-assigned-spin resonances in the $^{22}$Mg spectrum. Finally, Fig.\ \ref{fig:xsec_comp_results} shows a calculation with the Hauser-Feshbach code SMARAGD \cite{smaragd1, smaragd2}, which includes the cross sections leading towards the ground state and first five excited states of $^{21}$Na, including the unbound 2.798~MeV state. The overall agreement with our data is excellent. 

\begin{figure}[h]
	\centering
\includegraphics[width=0.5\textwidth]{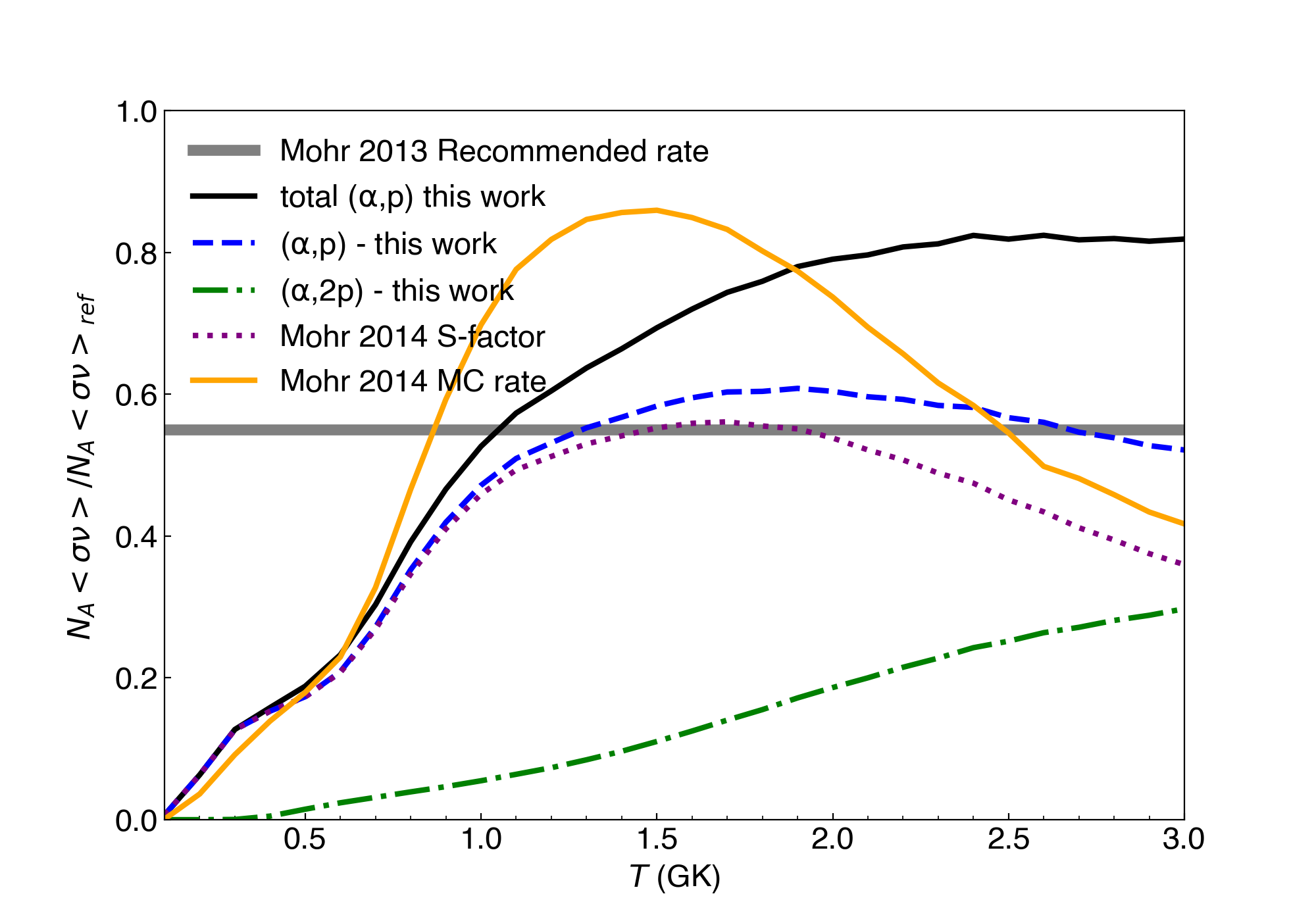}
	\caption{Comparison of the reaction rates of this work and Mohr {\it{et al.}} 2014 work \cite{mohr2014} to the reference reaction rate $N_{A}<\sigma v>_{ref}$ determined in Ref.~\cite{mohr2013}. The flat line at $0.55~N_{A}<\sigma v>_{ref}$ is the recommended reaction rate from Ref.~\cite{mohr2013}. The total ($\alpha,p$) reaction rate of this work is larger by a factor of 1.5 from the recommended rate of Ref.~\cite{mohr2013} due to the contributions of the $^{21}$Na unbound states. }
	\label{fig:RR_comparison}
\end{figure}

The impact of our results on the reaction rate is illustrated in Fig.~\ref{fig:RR_comparison}.  We calculate reaction rates using the unfolded cross sections from this work shown in Figs. \ref{xsections}--\ref{fig:xsec_comp_results} for $E_{cm} > 2.5$~MeV. For $E_{cm}< 2.5$~MeV, we adopt cross sections based on the resonant properties given in Table I of Mohr {\it{et al.}}~\cite{mohr2014}, where we are in agreement and our statistical precision is poor.  Following the presentation in \cite{mohr2014}, we plot the ratio of our reaction rates to the reference rate from Ref. \cite{mohr2013}.
At peak X-ray burst temperatures $T=2.5$~GK, the $^{18}$Ne$(\alpha,p)^{21}$Na reaction rate from a previous Monte Carlo analysis in Ref. \cite{mohr2014} (yellow, solid curve in Fig.~\ref{fig:RR_comparison}) agrees with the recommended reaction rate from Ref.~\cite{mohr2013}, which was given as 55$\%$ of the reference rate (flat grey line in Fig.~\ref{fig:RR_comparison}). We find the experimental reaction rate to be a factor of 1.5 larger at $T=2.5$~GK (black, solid curve Fig.~\ref{fig:RR_comparison}). The increase comes primarily from reactions proceeding through proton unbound states in $^{21}$Na that subsequently decay by proton emission to $^{20}$Ne, as can be seen in the comparison of the ($\alpha,p$) (blue, dashed curve Fig.~\ref{fig:RR_comparison}) to the ($\alpha,2p$) (green, dashed-dotted curve) contributions to the reaction rate. The resonance properties in $^{22}$Mg used in previous estimates of the reaction rate come almost exclusively from 2-neutron stripping reactions.  It is not surprising that broad proton-unbound states may have been missed that would account for the enhanced contribution from $^{18}$Ne$(\alpha,2p)^{20}$Ne that we observe. In most astrophysical scenarios the $^{18}$Ne$(\alpha,2p)^{20}$Ne branch is likely to have a negligible effect due to the high rate of the $^{20}$Ne($p$,$\gamma$)$^{21}$Na reaction; however, it is advisable to include the $^{18}$Ne$(\alpha,2p)^{20}$Ne reaction rate as a separate channel, particularly in models with low hydrogen abundance, such as may result from accretion from a helium-rich companion. 

Finally, it is worth noting that the Monte Carlo estimate of the $^{18}$Ne$(\alpha,p)^{21}$Na reaction rate in Ref.~\cite{mohr2014} (yellow curve in Fig.~\ref{fig:RR_comparison}) is significantly higher for $T \approx 1-2$~GK than that obtained using the resonance parameters (Table I of \cite{mohr2014}) or a direct integration using the recommended S-factor (Fig. 2 of \cite{mohr2014}) (purple, dotted curve Fig.~\ref{fig:RR_comparison}). Since we adopt the resonance parameterization for $E_{cm}< 2.5$~MeV in calculating the reaction rate, our reaction rate may similarly underestimate the reaction rate in this intermediate temperature range.

\section{\label{sec:level1-Conclusions} Conclusions}

Our experiment determined cross sections for the $^{18}$Ne($\alpha,p$)$^{21}$Na reaction between about 2.5-4~MeV in the center of mass using the active-target detector system ANASEN, which was augmented by an ion-chamber component, allowing for a higher detection sensitivity without limiting the usable beam rate. The present experiment is the first measurement of an ($\alpha,p$) reaction on a radioactive beam with the ANASEN active-target detector, determining cross sections as low as 5 mbarn and in the center of the Gamow-window for a temperature $T = 2$~GK. The partial cross sections for the population of the $^{21}$Na ground state are consistent with those derived from experiments on the time-reverse reaction by Sihna {\it et al.} \cite{ANL} and Salter {\it et al.} \cite{salter}. The total cross section is significantly lower than the one found in previous experiments by Bradfield-Smith {\it et al.} \cite{bradfield} and Groombridge {\it et al.} \cite{groom}. 

Compared to the predictions of Mohr {\it et al.} \cite{mohr2013, mohr2014}, which are based on indirectly determined resonance parameters, our cross sections are larger at higher reaction energies, demonstrating the contributions from the proton-unbound states of $^{21}$Na. While in most astrophysical scenarios the $^{18}$Ne$(\alpha,2p)^{20}$Ne branch is likely to have a negligible effect due to the high rate of the $^{20}$Ne($p$,$\gamma$)$^{21}$Na reaction, the $^{18}$Ne$(\alpha,2p)^{20}$Ne reaction rate may have a separate impact on models with low hydrogen abundance, such as those from accretion from a helium-rich companion. Finally, the cross sections of this work are consistent with the predictions of a Hauser-Feshbach calculation. Although at low energies the reaction proceeds mostly through individual resonances, which could not be resolved by our experiment, the overall agreement with Hauser-Feshbach calculations seems to justify their use as an approximation to determine the thermal reaction rate until more experimental data is available for the direct $^{18}$Ne($\alpha,p$)$^{21}$Na. 

\section{\label{sec:level1-Acknowledgments} Acknowledgments}

This work was partially supported by the National Science Foundation under Grants No.~PHY-1712953 and PHY-2012522, and partially supported by the U.S. Department of Energy, Office of Science under Grants No.~DE-FG02-96ER40978 and No.~DE-FG02-93ER40773. T.R. is partially supported by the “ChETEC” COST Action (CA16117). M.A. was supported for the writing of this manuscript (LLNL-JRNL-821361) by the U.S. Department of Energy by Lawrence Livermore National Laboratory under Contract DE-AC52-07NA27344. 

%

\end{document}